\documentclass[english]{czjphys}

\usepackage[reqno,centertags,intlimits]{amsmath}
\usepackage{amsfonts}
\usepackage{amssymb}
\usepackage{babel}
\usepackage{graphicx}

\usepackage{DSRFTD}

\usepackage{hyperref}
\begin{document}

\title
{Asymptotic directional structure of radiation\\ for fields of algebraic type D%
\footnote{Dedicated to Prof. Ji\v{r}\'{\i} Hor\'{a}\v{c}ek on the occasion of his 60th birthday}}

\authori{Pavel Krtou\v{s} and Ji\v{r}\'{\i} Podolsk\'y}
\addressi{%
  Institute of Theoretical Physics,
  Faculty of Mathematics and Physics, Charles University,\\
  V Hole\v{s}ovi\v{c}k\'{a}ch 2, 180 00 Praha 8, Czech Republic
  }
\authorii{}     \addressii{}
\authoriii{}    \addressiii{}
\authoriv{}     \addressiv{}
\authorv{}      \addressv{}
\authorvi{}     \addressvi{}

\headauthor{Pavel Krtou\v{s} and Ji\v{r}\'{\i} Podolsk\'y}            % page heading on the even pages
\headtitle{Asymptotic directional structure of radiation for algebraically simple fields}             % page heading on the odd pages
\lastevenhead{Pavel Krtou\v{s} and Ji\v{r}\'{\i} Podolsk\'y: Asymptotic directional structure \dots} % p. h. on the last page if even



%%%%%%%%%%%%%% FOR EDITORIAL USE ONLY!!! %%%%%%%%%%%%%%%
\refnum{}%\total{}\type{}
\daterec{}
\issuenumber{2, 119--138}  \year{2005}
\setcounter{page}{1}
%\firstpage{1}
%\lastpage{000}
%\makefirsttitle
%%%%%%%%%%%%%%%%%%%%%%%%%%%%%%%%%%%%%%%%%%%%%%%%%%%%%%%%

\maketitle

\begin{abstract}
The directional behavior of dominant components of algebraically 
special spin-$s$ fields near a spacelike, timelike or null 
conformal infinity is studied. By extending our previous general 
investigations we concentrate on fields which admit a pair of 
equivalent algebraically special null directions, such as 
the Petrov type D gravitational fields or algebraically general 
electromagnetic fields. We introduce and discuss a canonical 
choice of the reference tetrad near infinity in all possible 
situations, and we present the corresponding asymptotic 
directional structures using the most natural parametrizations.
\end{abstract}

%\newpage

\section{Introduction}
\vspace{-1ex}
${\mathbf{}}$

In the series of papers \cite{KrtousPodolsky:2003,PodolskyOrtaggioKrtous:2003,KrtousPodolskyBicak:2003,KrtousPodolsky:2004a} 
we analyzed the asymptotic directional properties of electromagnetic 
and gravitational fields in spacetimes with a nonvanishing cosmological 
constant $\Lambda$. It had been known for a long time \cite{Penrose:1964,Penrose:1965,Penrose:1967} that 
--- contrary to the asymptotically flat spacetimes --- the dominant 
(radiative) component of the fields is not unique since it 
depends substantially on the direction along which a null geodesic 
approaches a given point at conformal infinity $\scri$. 
We demonstrated that, somewhat surprisingly, such directional 
structure of radiation can be  described in closed 
explicit form. It has a universal character that is 
essentially determined by the algebraic type of the field, i.e., 
by the specific local degeneracy and orientation of the principal null directions.

Our results were summarized and thoroughly discussed in the recent 
topical review \cite{KrtousPodolsky:review}. They apply not only to electromagnetic 
or gravitational fields but to any field of spin $s$. 
In addition, the expression representing the directional behavior of radiation 
can be written in a unified form which covers all three possibilities 
$\Lambda>0$, $\Lambda<0$ or $\Lambda=0$, corresponding  to a spacelike, 
timelike or null character of~$\scri$, respectively.

This paper further elaborates and supplements some aspects of our work  
reviewed in \cite{KrtousPodolsky:review}. It extends possible 
definitions of the canonical reference tetrad for the algebraically simple fields ---
those which admit an equivalent pair of distinct (degenerate) principal null directions.
We systematically describe the most natural choices of the reference tetrad for all 
possible algebraic structures of type D fields, and for any value of $\Lambda$.

The notation used in the present paper is the same as in \cite{KrtousPodolsky:review}; 
in fact, we will frequently employ the material already described and derived there. 
For brevity, we will refer directly to the equations and sections of the review
\cite{KrtousPodolsky:review} by prefixing the letter `R' in
front of the reference: equation (\revref{1.1}), section \revref{2.1}, etc.

\section{Summary of general results}

We wish to study the behavior of radiative component of  fields near 
conformal infinity ${\scri}$. An overview of the concept of conformal infinity can
be found in textbooks (e.g., \cite{PenroseRindler:book};
our notation is described in section \revref{2} of the work \cite{KrtousPodolsky:review}).

Let us only recall that it is possible to define a normalized vector ${\norm}$ normal to the conformal
infinity. The causal character of the infinity --- spacelike, null, or timelike ---
is given by the sign of the square of this vector, ${\nsgn=\norm\spr\norm=-1,0,+1}$
(see also figure \ref{fig:scri}).
Here and in the following, the dot `$\spr$' denotes the scalar product
defined using the spacetime metric ${\mtrc}$. Typically, the causal character of $\scri$ is
correlated with the sign of the cosmological constant, ${\nsgn=-\sign\Lambda}$
(see section \revref{2.2} for details).

\subsection{Null tetrads}

To study various components of the fields we
introduce suitable orthonormal, and
associated with them null tetrads. We denote the vectors of
an \defterm{orthonormal tetrad}  as ${\tG,\,\qG,\,\rG,\,\sG}$,
where $\tG$ is a future oriented unit timelike vector.
With this tetrad we associate a \defterm{null tetrad}
of null vectors ${\kG,\,\lG,\,\mG,\,\bG}$ by
\begin{equation}\label{NormNullTetr}
  \kG = \textstyle{\frac1{\sqrt{2}}} (\tG+\qG)\comma
  \lG = \textstyle{\frac1{\sqrt{2}}} (\tG-\qG)\comma
  \mG = \textstyle{\frac1{\sqrt{2}}} (\rG-i\,\sG)\comma
  \bG = \textstyle{\frac1{\sqrt{2}}} (\rG+i\,\sG)\period
\end{equation}
The normalization conditions are
\begin{equation}\label{TetrNorm}
-\tG\spr\tG=\qG\spr\qG=\rG\spr\rG=\sG\spr\sG=1\comma
-\kG\spr\lG=\mG\spr\bG=1\commae
\end{equation}
respectively, with all other scalar products being zero.

The crucial tetrad in our study is the \defterm{interpretation tetrad}
${\kI,\,\lI,\,\mI,\,\bI}$. It is a tetrad which is parallelly transported along
a null geodesic ${\geod(\afp)}$, the vector ${\kI}$ being tangent to the geodesic.
With respect to this tetrad we define the radiative
component of the field. The precise definition and description of asymptotic behavior of the
interpretation tetrad was thoroughly presented in sections \revref{3.3} and \revref{3.4}
where more details can be found.

\begin{figure}[t]
\begin{center}
\includegraphics{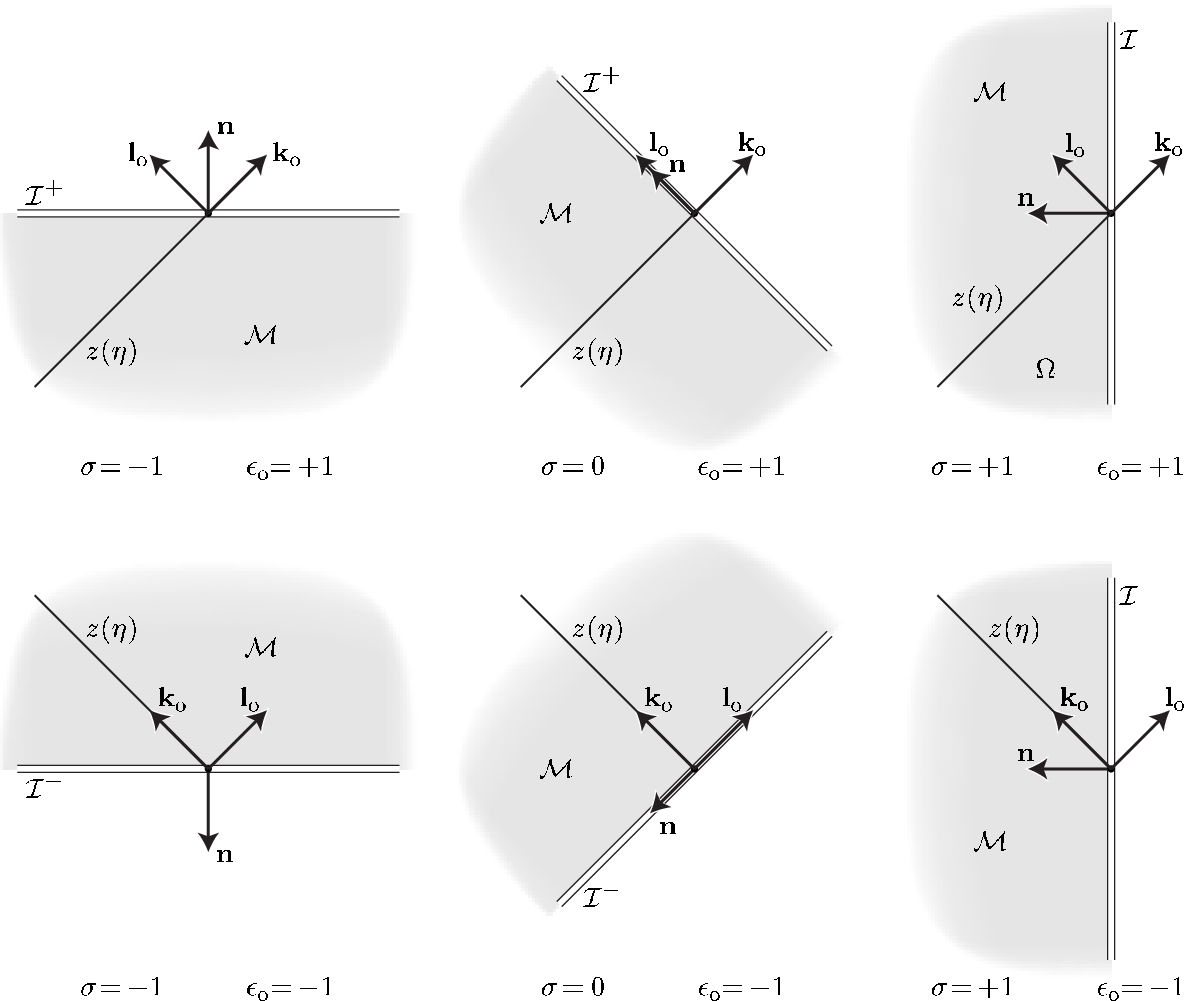}
\end{center}
\caption{\label{fig:scri}%
The reference tetrad adjusted to conformal
infinity $\scri$ of various character, determined by $\nsgn$
which is a norm of the vector ${\norm}$ normal to ${\scri}$.
If the vector $\kO$ is oriented along an outgoing direction
(outward from the physical spacetime ${\mfld}$) we have ${\EPS_\refT=+1}$,
if it is  ingoing 
(oriented inward to ${\mfld}$) then ${\EPS_\refT=-1}$.}
\end{figure}

Here, we are going to concentrate on the \defterm{reference tetrad} ${\kO,\,\lO,\,\mO,\,\bO}$.
It is a tetrad conveniently defined near the conformal infinity.
It serves as the reference frame for parametrization of directions near ${\scri}$. It can be
defined using special features of the spacetime geometry (e.g., the Killing vectors,
direction toward sources, etc.). Alternatively, it can be adapted to
the studied fields --- namely, it can be \vague{aligned} with
algebraically special directions of the fields under consideration.
A general situation was discussed in sections
\revref{5.4} and \revref{5.5}. In the present work we will
offer other possible privileged definitions of the reference tetrad for algebraically
simple fields of type D.

Following \cite{KrtousPodolsky:review} we require that
the reference tetrad is \defterm{adjusted to conformal infinity},
i.e., that the vectors $\kO$ and $\lO$  satisfy the relation
\begin{equation}\label{AdjustedRefer}
   \norm = \EPS_{\refT}\textstyle{\frac{1}{\sqrt{2}}}
   (-\nsgn\kO+\lO)\commae
\end{equation}
where the sign ${\EPS_\refT=\pm1}$ indicates the outgoing/ingoing orientation of the vector ${\kO}$
with respect to ${\scri}$ (see also below). This adjustment condition guarantees that the
vectors $\kO$ and $\lO$ are colinear with the normal ${\norm}$ to ${\scri}$, and normalized such that
\begin{equation}\label{AdjustedReferNorm}
  \norm =
\begin{cases}
  \;\EPS_\refT\,\tO           \qquad&\text{for a spacelike infinity (${\nsgn=-1}$)}\commae\\[1pt]
  \;-\EPS_\refT\,\qO          \qquad&\text{for a timelike infinity (${\nsgn=+1}$)}\commae\\[1pt]
  \;\EPS_\refT\,\lO/\sqrt{2}  \qquad&\text{for a null infinity (${\nsgn=0}$)}\period\\
\end{cases}
\end{equation}
All possible orientations of the reference tetrad with respect to $\scri$ are shown in figure \ref{fig:scri}.

The normalization and adjustment conditions do not fix the reference tetrad
uniquely. Additional necessary conditions --- the alignment with the algebraically special
directions --- will be specified in  section \ref{sc:TypeD}.

\subsection{Parametrization of null directions}
\label{ssc:nulldir}

In the following it will be necessary to parametrize
a general null direction ${\kG}$ near~${\scri}$.
This can be done with respect to the reference tetrad by a complex \defterm{directional parameter ${R}$}:
\begin{equation}\label{Rmeaning}
  \kG \propto\kO + \bar{\R}\, \mO + \R\, \bO + \R\bar{\R}\, \lO \period
\end{equation}
The value ${\R=\infty}$ is also permitted --- it corresponds to
$\kG$ oriented along $\lO$.

In addition, we introduce the \defterm{orientation parameter ${\EPS}$} which indicates whether
the null direction ${\kG}$ is an outgoing direction (pointing outside
the spacetime), ${\EPS=+1}$, or if it is an ingoing direction (pointing inside
the spacetime), ${\EPS=-1}$.

When the infinity ${\scri}$ has a \emph{spacelike} character it is also possible to parametrize
the null direction ${\kG}$ using \defterm{spherical angles} which specify its normalized spatial projection into
${\scri}$. We define the angles ${\THT,\,\PHI}$ by
\begin{equation}\label{THTPHIdef}
  \qG = \cos\THT\;\qO + \sin\THT\,(\cos\PHI\;\rO + \sin\PHI\;\sO)\commae
\end{equation}
where ${\qG}$ is the unit vector pointing into the spatial (${\qG\spr\norm=0}$)
direction given by ${\kG}$, see (\revref{5.5}).
The complex parameter ${R}$ of \eqref{Rmeaning} is actually
a stereographic representation of the spatial direction ${\qG}$:
\begin{equation}\label{RTHTPHIrel}
  R = \tan\frac{\THT}{2}\,\exp(-i\PHI)\period
\end{equation}

Near a \emph{timelike} infinity ${\scri}$ we can analogously describe null direction ${\kG}$ by
\defterm{pseudospherical parameters ${\PSI,\,\PHI}$} of its projection into
${\scri}$. If we label the normalized projection of ${\kG}$ by ${\tG}$,
cf.~(\revref{5.8}), ${\PSI}$ and ${\PHI}$ are given by
\begin{equation}\label{PSIPHIdef}
  \tG = \cosh\PSI\;\tO + \sinh\PSI\,(\cos\PHI\;\rO + \sin\PHI\;\sO) \period
\end{equation}
These parameters have to be supplemented by the orientation ${\EPS}$ of ${\kG}$ with respect to
${\scri}$. The parameter ${R}$ is the pseudostereographic representation of ${\tG}$:
\begin{equation}\label{RPSIPHIrel}
  \R = \tanh^{\EPS\EPS_\refT}\!\frac\PSI2\;\exp(-i\PHI)\period
\end{equation}

In fact, we can introduce both these parametrizations simultaneously, independently of the causal character
of the infinity, just with respect to the reference tetrad
--- the angles ${\THT,\,\PHI}$ using a projection onto the 3-space
orthogonal to the time vector ${\tO}$ of the reference tetrad, and the parameters
${\PSI,\,\PHI}$ using a projection to the 2+1-space orthogonal to ${\qO}$. In such
a case they are related by expressions
\begin{equation}\label{THTPSIrel}
\tanh\PSI=\sin\THT\comma
\sinh\PSI=\tan\THT\comma
\cosh\PSI=\cos^{\!-\!1}\!\THT\comma
{\textstyle \tanh\frac\PSI2=\tan\frac\THT2}\period
\end{equation}

If the infinity has a timelike character, all future oriented null directions
at one point at ${\scri}$
naturally split into two families of outgoing and ingoing directions.
The directions of each of these families form a hemisphere which can be projected
onto a unit circle parametrized by ${\RHO}$ and ${\PHI}$, such that
\begin{equation}\label{RHOdef}
\RHO=\tanh\PSI=\sin\THT\commae
\end{equation}
see also figure \revref{3}.

\subsection{Asymptotic directional structure of radiation}

The main result of the paper \cite{KrtousPodolsky:review} is  derivation of the
explicit dependence of the radiative component of the field on a
direction along which the infinity is approached --- we call
this dependence the \defterm{asymptotic directional structure of radiation}.

In \cite{KrtousPodolsky:review} we  investigated a general spin-${s}$ field, and in more detail
gravitational (${s=2}$) and electromagnetic (${s=1}$) fields.
These fields can be characterized by ${2s+1}$ complex components
${\fieldP{}{j}}$, ${j=0,\dots,2s}$,
evaluated with respect to a null tetrad. Relation of these components
to a spinor representation of the fields, and their
transformation properties can be found in \revref{4.1} and appendix \revref{B}.
The components of gravitational and electromagnetic fields are traditionally
called ${\WTP{}{j}}$, ${j=0,\dots,4}$, and ${\EMP{}{j}}$, ${j=0,1,2}$,
respectively --- see \cite{Stephanietal:book} or equations (\revref{4.1}) and
(\revref{4.2}).

To study the asymptotic behavior we evaluated the field with
respect to the interpretational tetrad --- the tetrad which is parallelly
transported along a null geodesic ${\geod(\afp)}$.
It turned out that these field components satisfy
the standard \defterm{peeling-off} property, namely that they exhibit a different fall-off in ${\afp}$
when approaching ${\scri}$, ${\afp}$~being the affine parameter of the geodesic. The leading
component of the field is the component ${\fieldP{\intT}{\!2\!s}}$ with the fall-off of
order ${\afp^{-1}}$, and we call it the \defterm{radiative component}.
In sections \revref{4.3} and \revref{4.4} we found that the radiative field component
depends on the direction ${\R}$ of the null geodesic along which a fixed point at the infinity
is approached. This directional structure is determined mainly by the algebraic structure of
the field, and it reads
\begin{equation}\label{DSRspin}
\fieldP{\intT}{\!2\!s} \lteq \frac{1}{\afp} \,\,\EPS_\refT^s\fieldP{\refT}{\!2\!s*}\,
  \frac{\bigl(1-\nsgn\R_1\bar{\R}\bigr)\bigl(1-\nsgn\R_2\bar{\R}\bigr)\ldots\bigl(1-\nsgn\R_{2\!s}\bar{\R}\bigr)}
  {\bigl(1-\nsgn \R\bar{\R}\bigr)^s}  \period
\end{equation}
The complex constants ${R_1,\dots,R_{2\!s}}$ represent the \defterm{principal null
directions} ${\PND{1},\dots,\PND{2\!s}}$ of the spin-$s$ field,
the sign ${\nsgn=\pm1,0}$ specifies the causal character of the conformal infinity,
${\EPS_\refT}$ denotes orientation of the
reference tetrad, and ${\fieldP{\refT}{2\!s*}}$ is a constant normalization factor of
the field evaluated with respect to the reference tetrad,
${\fieldP{\refT}{2\!s}\lteq \fieldP{\refT}{2\!s*}\afp^{-s-1}}$ (cf. section
\revref{4.4}).

The principal null directions (PNDs)
are special directions along which some of the field
components vanish --- see \cite{PenroseRindler:book,Stephanietal:book} or
section \revref{4.2} for a precise definition. The field of spin ${s}$ has ${2s}$
PNDs. However, these can be degenerate, and this degeneracy
(or, more generally, mutual relations of all the PNDs) is called
the \defterm{algebraic structure} of the field. Distinct PNDs are also called
\defterm{algebraically special directions} of the field.
The classification according to the degeneracy of PNDs
for a gravitational field is the well-known Petrov classification.

\section{Fields of type {D}}
\label{sc:TypeD}

In this paper we wish to discuss the situation when the field has two distinct and
equivalent algebraically special directions.
This may occur only for fields of an \emph{integer} spin, ${s\!\in\!\naturaln}$,
with PNDs having the degeneracy
\begin{equation}\label{degPNDs}
\PND{1}=\dots=\PND{s}\quad\text{and}\quad
\PND{s+1}=\dots=\PND{2\!s}\period
\end{equation}
The directional structure \eqref{DSRspin} of such a field takes the form
\begin{equation}\label{TypeDDSRspin}
\fieldP{\intT}{\!2\!s} \lteq \frac{1}{\afp} \,\,\EPS_\refT^s\fieldP{\refT}{\!2\!s*}\,
  \frac{\bigl(1-\nsgn\R_1\bar{\R}\bigr)^s\bigl(1-\nsgn\R_{2\!s}\bar{\R}\bigr)^s}
  {\bigl(1-\nsgn \R\bar{\R}\bigr)^s}  \commae
\end{equation}
with the constants ${R_1}$ and ${R_{2\!s}}$ parametrizing the two distinct PNDs. %${\PND1}$ and ${\PND{2\!s}}$.
The directional dependence of the magnitude of a gravitational type {D} field
and of an algebraically general electromagnetic field are thus quite similar.
This similarity is even closer if we recall that the square of the electromagnetic
component $\EMP{\intT}{2}$ is proportional to the magnitude of the Poynting
vector with respect to the interpretation tetrad,
${\abs{\EMS_\intT}\lteq\frac1{4\pi}\abs{\EMP{\intT}{2}}^2}$.
Indeed, we have
\begin{gather}
\abs{\WTP{\intT}{4}} \lteq
  \frac{1}{\abs{\afp}} \,\abs{\WTP{\refT}{4*}}\,
  \frac{\abs{1-\nsgn\R_1\bar{\R}}^2\abs{1-\nsgn\R_4\bar{\R}}^2}
  {\abs{1-\nsgn \R\bar{\R}}^2}
  \commae  \label{TypeDDSRgr}\\
4\pi \abs{\EMS_\intT} \lteq \abs{\EMP{\intT}{2}}^2 \lteq
  \frac{1}{\afp^2} \,\abs{\EMP{\refT}{2*}}^2\,
  \frac{\abs{1-\nsgn\R_1\bar{\R}}^2\abs{1-\nsgn\R_2\bar{\R}}^2}
  {\abs{1-\nsgn \R\bar{\R}}^2}
  \period  \label{TypeDDSRem}
\end{gather}

As discussed in section \revref{4.5}, the form of the directional
structure \eqref{DSRspin} depends on the choice of the normalization
factor ${\fieldP{\refT}{\!2\!s*}}$. Other choices can sometimes be more convenient,
in particular if the factor ${\fieldP{\refT}{\!2\!s*}}$ vanishes which happens when one of
the PNDs points along the direction ${\lO}$ of the reference tetrad. For the type D fields
there exists a more natural \vague{symmetric} choice of normalization of the directional structure of
radiation which is guaranteed to be non-degenerate. For these fields we can
define a \defterm{canonical field component}, namely, the only nonvanishing component
with respect of the null tetrad associated with the PNDs \eqref{degPNDs}.

Having two distinct algebraic directions ${\PND{1}}$ and ${\PND{2\!s}}$ we can
define \defterm{algebraically special null tetrad\, ${\kS,\,\lS,\,\mS,\,\bS}$}
(and associated orthonormal tetrad ${\tS,\,\qS,\,\rS,\,\sS}$) by requiring that
${\kS,\,\lS}$ are \emph{proportional} to the PNDs and future oriented, and that
the spatial vector ${\sS}$ is \emph{tangent} to ${\scri}$,
\begin{equation}\label{algspectetr}
\kS\propto\PND{1}\comma\lS\propto\PND{2\!s}\comma\sS\spr\norm=0\period
\end{equation}
For PNDs which are not tangent to the conformal infinity the normalization of null vectors
${\kS,\,\lS}$ can be fixed by condition
\begin{equation}\label{klspcnorm}
\EPS_1\,\kS\spr\norm = \EPS_{2\!s}\,\lS\spr\norm\commae
\end{equation}
where ${\EPS_1,\,\EPS_{2\!s}=\pm1}$ parametrize orientations of the PNDs with respect to ${\scri}$.
The special case of PNDs tangent to ${\scri}$ will be discussed below.

Using the definition of PNDs (see section \revref{4.2}) we find that the field components
with respect to the algebraically special tetrad
have very special form --- only the component ${\fieldP{\spcT}{s}}$ is nonvanishing
(${\WTP{\spcT}{2}}$ for gravitational and ${\EMP{\spcT}{1}}$ for electromagnetic fields).
This component  is, in fact, independent
of the choice of the spatial vectors ${\rS,\,\sS}$, and thus it does not depend on 
the normal vector ${\norm}$ which we used
in the definition~\eqref{algspectetr}. It also does not depend on the normalization \eqref{klspcnorm},
provided that the normalization \eqref{TetrNorm} is satisfied.
We will use the privileged component ${\fieldP{\spcT}{s}}$ for the normalization of the directional structure of radiation.
However, the algebraically special tetrad is \emph{not} adjusted to the infinity
(cf. condition~\eqref{AdjustedRefer}) since ${\kS}$ and ${\lS}$ are not in general colinear with ${\norm}$
and thus it cannot  be used as a reference tetrad.

Nevertheless, we can define privileged reference tetrad which is
\vague{somehow aligned} with the algebraically special tetrad, and which  shares some
of the symmetries of the geometric situation. We will always assume that the reference tetrad
satisfies the normalization and adjustment conditions \eqref{TetrNorm}, \eqref{AdjustedRefer},
and we set ${\sO=\sS}$. This is, however, still not sufficient to fix the reference tetrad completely.
The remaining necessary condition cannot be prescribed in general
--- we have to discuss separately several possible cases, depending on the character of
the infinity ${\scri}$, and on the orientation of the PNDs with respect to ${\scri}$.

Below we will define the reference tetrad for all possible cases.
We will present the relation between the component ${\fieldP{\refT}{\!2\!s}}$
and the canonical component ${\fieldP{\spcT}{s}}$,
which can be substituted into the directional
structure~\eqref{TypeDDSRspin}. Finally, we will rewrite the results in terms of the angular
variables introduced with respect to the reference tetrad.

\subsection{Spacelike ${\scri}$}
\label{ssc:spacelike}

\begin{figure}[t]
\begin{center}
\includegraphics{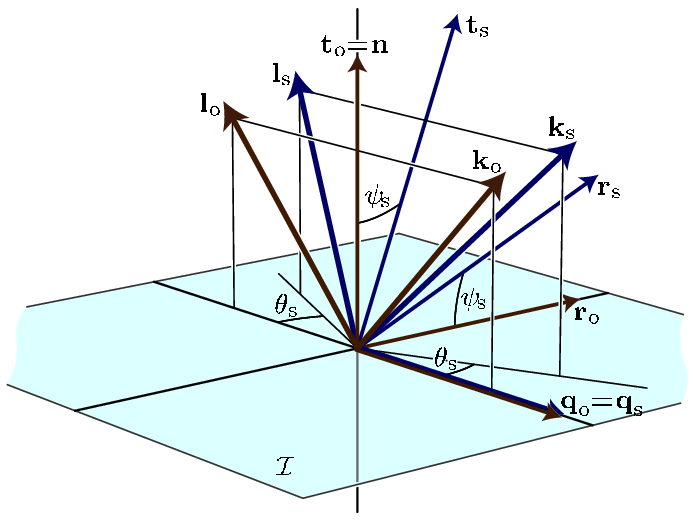}
\end{center}
\caption{\label{fig:ast-S}%
Algebraically special and reference tetrads at a spacelike infinity. 
Vectors ${\kO,\,\lO}$ (${\kS,\,\lS}$, respectively) of the null tetrad, and
$\tO$, $\qO$, $\rO$ ($\tS$, $\qS$, $\rS$) of 
the orthonormal reference (algebraically special, respectively)
tetrad are shown; the direction ${\sO=\sS}$ tangent to $\scri$ and orthogonal to PNDs is hidden.
The vectors $\kS$, $\lS$ are aligned with algebraically special directions
(degenerate PNDs), $\tO$ is normal to infinity $\scri$, and $\qO$, $\rO$ 
are tangent to~$\scri$. The relation of both the tetrads is parametrized
by the angle $\THTS$ between $\qO$ and the projection of $\kS$ onto $\scri$.
The special tetrad can be obtained from the reference tetrad by a boost
in ${\tO\textdash\rO}$ plane with rapidity parameter $\PSIS$ given by
 ${\sinh\PSIS=\tan\THTS}$, cf.\ \eqref{RefSpcSpacelike}.} 
\end{figure}

We begin with a spacelike conformal infinity, ${\nsgn=-1}$.
In this case the two distinct future oriented algebraically special directions are either both
ingoing or both outgoing, and we set the orientation ${\EPS_\refT}$ of the reference
tetrad accordingly.
We define the reference tetrad by conditions
\begin{equation}\label{spctetrdefspacelike}
\qO=\qS\comma \sO=\sS\comma \EPS_\refT=\EPS_1=\EPS_{2\!s}\commae
\end{equation}
and by adjustment condition \eqref{AdjustedReferNorm}. It follows that the algebraically
special directions ${\kS}$ and ${\lS}$ are parametrized with respect to the
reference tetrad by a single parameter~${\THTS}$ as
\begin{equation}
\begin{aligned}
\kS&={\textstyle\frac1{\sqrt2}}\cos^{\!-1\!}\THTS\,\bigl(\tO+\cos\THTS\,\qO+\sin\THTS\,\rO\bigr)\commae\\
\lS&={\textstyle\frac1{\sqrt2}}\cos^{\!-1\!}\THTS\,\bigl(\tO-\cos\THTS\,\qO+\sin\THTS\,\rO\bigr)\period
\end{aligned}
\end{equation}
The algebraically special and reference tetrads are thus related by
\begin{equation}\label{RefSpcSpacelike}
\begin{gathered}
\tS=\cos^{\!-1\!}\THTS\,\tO+\tan\THTS\,\rO\comma \qS=\qO\comma
\rS=\cos^{\!-1\!}\THTS\,\rO+\tan\THTS\,\tO\comma \sS=\sO\commae
\end{gathered}
\end{equation}
which is actually a boost in ${\tO\textdash\rO}$ plane with rapidity parameter ${\PSIS}$
related to $\THTS$ by \eqref{THTPSIrel},  see figure \ref{fig:ast-S}.

Inspecting the spatial projections of ${\kS}$ and ${\lS}$ onto conformal infinity ${\scri}$,
we find that their angular coordinates with respect to the reference tetrad are
${\THT_1=\THTS}$, ${\PHI_1=0}$, and ${\THT_{2\!s}=\pi-\THTS}$, ${\PHI_{2\!s}=0}$, respectively.
It means that the complex parameters ${R_1}$ and ${R_{2\!s}}$ of both these algebraic
special directions are
\begin{equation}\label{PNDRspacelike}
\R_1=\tan\frac\THTS2\comma
\R_{2\!s}=\cot\frac\THTS2\period
\end{equation}

Straightforward calculation shows that the transformation
\eqref{RefSpcSpacelike} from the algebraically special to the reference
tetrad can be decomposed into boost (\revref{3.5}), subsequent null rotation with ${\kG}$~fixed (\revref{3.4}),
and  null rotation with ${\lG}$~fixed (\revref{3.3}), given by the parameters
${B\!=\!2(1+\cos^{\!-1\!}\THTS)^{-1}}$, ${L\!=\!-\frac12\tan\THTS}$, and
${K\!=\!-\tan(\THTS/2)}$. Applying these transformations to the field components
(relations (\revref{4.6}), (\revref{4.5}), and (\revref{4.4})) we easily find that
\begin{equation}\label{spcrefspinfieldspacelike}
\fieldP{\refT}{\!2\!s} = \binom{2s}{s}\bar L{}^s\,\fieldP{\spcT}{s} =
(-1)^s \frac{(2s)!}{2^s(s!)^2} \tan^s\THTS\,\fieldP{\spcT}{s}\period
\end{equation}
For gravitational and electromagnetic fields we can write explicit expressions
for all field components with respect to the reference tetrad as
\begin{gather}
\WTP{\refT}{0}=\WTP{\refT}{4}=\frac32\tan^2\!\THTS\,\WTP{\spcT}{2}\comma
\WTP{\refT}{1}=\WTP{\refT}{3}=-\frac32\frac{\tan\THTS}{\cos\THTS}\,\WTP{\spcT}{2}\comma
\WTP{\refT}{2}=\Bigl(1+\frac32\tan^2\!\THTS\Bigr)\WTP{\spcT}{2}\commae
\label{spcrefgrfieldspacelike}\\
\EMP{\refT}{0}=\EMP{\refT}{2}=-\tan\THTS\;\EMP{\spcT}{1}\comma
\EMP{\refT}{1}=\cos^{\!-\!1}\!\THTS\;\EMP{\spcT}{1}\period
\label{spcrefEMfieldspacelike}
\end{gather}

Following the discussion in section \revref{4.4} we assume asymptotic behavior (\revref{4.18}),
i.e., ${\fieldP{\refT}{j}\lteq\fieldP{\refT}{j*}\,\afp^{-s-1}}$
with constant coefficients ${\fieldP{\refT}{j*}}$. Similarly, we introduce
the coefficient ${\fieldP{\spcT}{s*}}$ by
${\fieldP{\spcT}{s}\lteq\fieldP{\spcT}{s*}\,\afp^{-s-1}}$.
Clearly, the relations \eqref{spcrefspinfieldspacelike}--\eqref{spcrefEMfieldspacelike}
hold also in their \vague{stared} forms.

\begin{figure}[t]
\begin{center}
\includegraphics{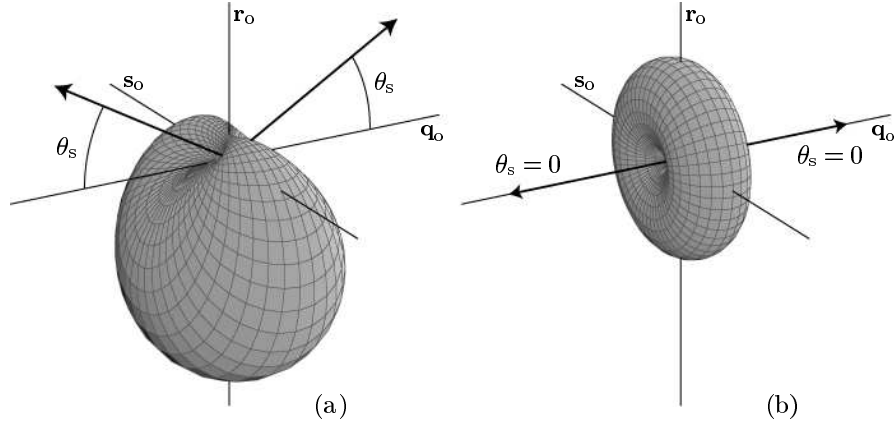}
\end{center}
\caption{\label{fig:dsr-S}%
Directional structure of radiation near a spacelike infinity. Directions in the 
diagrams correspond to spatial directions (projections onto ${\scri}$) 
of null geodesics along which the infinity is approached. The diagrams show 
the directional dependence of the magnitude of the radiative field component \eqref{TypeDDSRgrSpacelike} 
(or \eqref{TypeDDSRemSpacelike}). The arrows depict the directions which are spatially opposite to 
algebraically special directions (PNDs); the radiative component evaluated 
along the geodesics in these directions is asymptotically vanishing.
The diagram (a) shows a general orientation of algebraically special directions,
the diagram (b) corresponds to the case when both distinct PNDs are spatially opposite.}
\end{figure}

Substituting \eqref{PNDRspacelike}, the \vague{stared} version of
\eqref{spcrefspinfieldspacelike}, \eqref{RTHTPHIrel}, and ${\nsgn=+1}$ into
\eqref{TypeDDSRspin} we finally obtain the asymptotic directional structure
of radiation for type D fields%
\begin{equation}\label{TypeDDSRspinspacelike}
\fieldP{\intT}{\!2\!s} \lteq
  \frac{(-\EPS_\refT)^s}{\afp}\,
  \frac{(2s)!}{2^s(s!)^2}\,
  \fieldP{\spcT}{s*}\,
  \Bigl[\frac{\exp(i\PHI)}{\cos\THTS}\,
  \bigl(\sin\THT+\sin\THTS\cos\PHI-i\sin\THTS\cos\THT\sin\PHI\bigr)\Bigr]^s
  \period
\end{equation}
The null direction along which the field is measured is parametrized by angles ${\THT,\,\PHI}$,
the field itself is characterized by the normalization component ${\fieldP{\spcT}{s*}}$
and by the parameter ${\THTS}$ which encodes the directions of the
algebraically special directions with respect to a spacelike infinity~${\scri}$. As discussed
in section \revref{4.5}, only the magnitude of this radiative component has a physical meaning.
For the magnitude of the ${\WTP{\intT}{4}}$ component of the gravitational field
and for the Poynting vector of the electromagnetic field we thus obtain
\begin{gather}
\abs{\WTP{\intT}{4}} \lteq \frac{1}{\abs{\afp}}
\,\frac32\frac{\abs{\WTP{\spcT}{2*}}}{\cos^2\THTS}\,
  \bigl\lvert \sin\THT+\sin\THTS\cos\PHI-i \sin\THTS\cos\THT\sin\PHI\bigr\rvert^2\commae
  \label{TypeDDSRgrSpacelike}\\
4\pi\abs{\EMS_\intT}\lteq\abs{\EMP{\intT}{2}}^2\lteq
  \frac{1}{\afp^2} \,\frac{\abs{\EMP{\spcT}{1*}}^2}{\cos^2\THTS}\,
  \bigl\lvert \sin\THT+\sin\THTS\cos\PHI-i \sin\THTS\cos\THT\sin\PHI\bigr\rvert^2  \period
  \label{TypeDDSRemSpacelike}
\end{gather}
These are exactly the expressions for the asymptotic directional structure of
radiation as derived in \cite{BicakKrtous:2002} for test electromagnetic field of
accelerated charges in de~Sitter spacetime, and for gravitational filed and electromagnetic fields of the
$C$-metric spacetime with ${\Lambda>0}$, as presented in
\cite{KrtousPodolsky:2003}. This directional structure is illustrated in figure \ref{fig:dsr-S}.

\subsection{Timelike ${\scri}$ with non-tangent PNDs, ${\EPS_1\neq\EPS_{2\!s}}$}
\label{ssc:timelike1}

\begin{figure}[t]
\begin{center}
\includegraphics{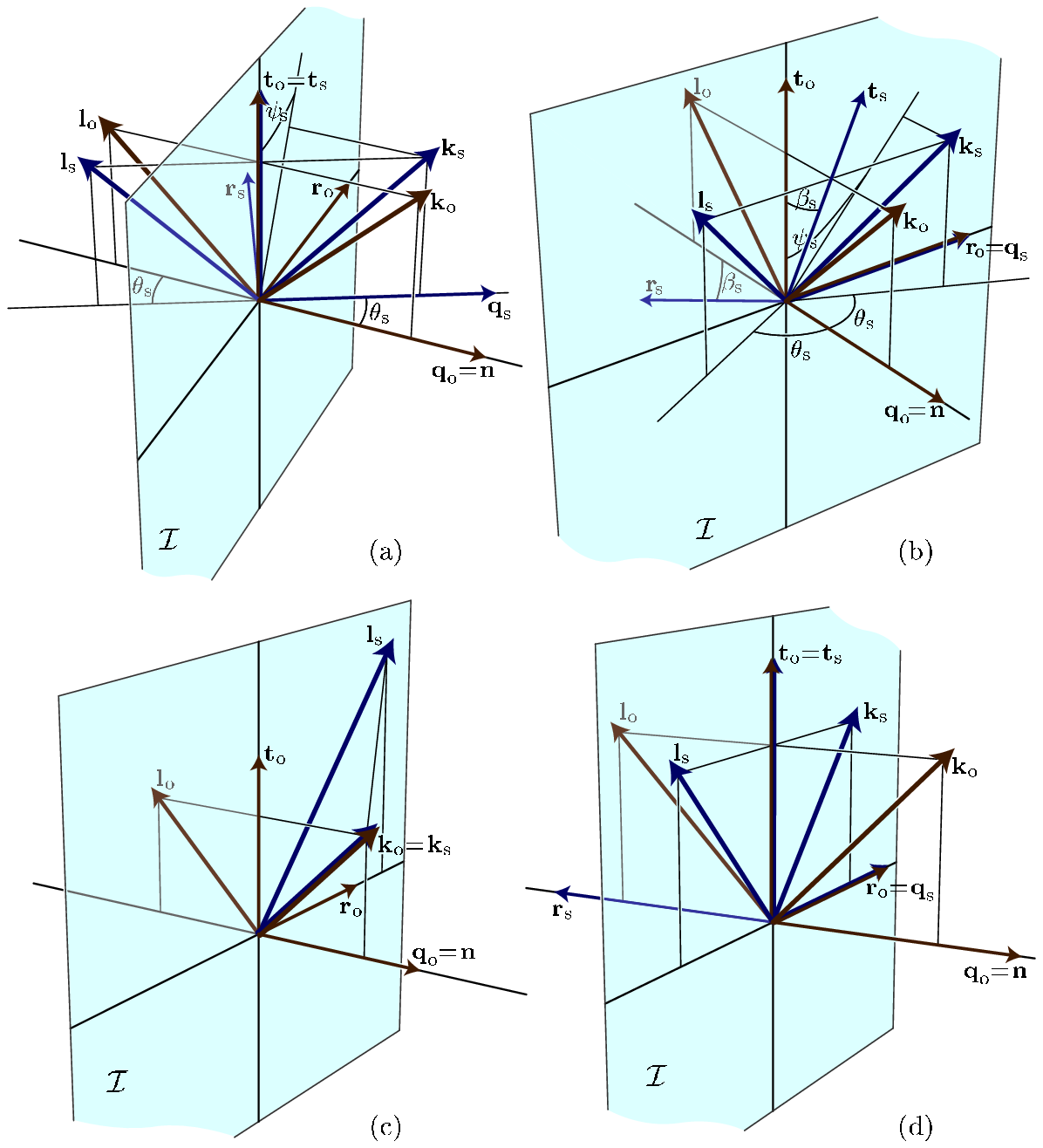}
\end{center}
\end{figure}
\begin{figure}[t]
\caption{\label{fig:ast-T}%
Algebraically special and reference tetrads near a timelike infinity.
Vectors ${\kO,\,\lO}$ (${\kS,\,\lS}$) of the null reference 
(algebraically special, respectively) tetrad are shown. The axes correspond 
to the timelike direction $\tO$ and the spatial directions $\qO$, $\rO$ 
of the reference tetrad. 
The direction ${\sO=\sS}$ tangent to $\scri$ and orthogonal to PNDs is hidden.
The vectors $\kS$, $\lS$ are aligned with the algebraically special directions
(degenerate PNDs), $\qO$ is normal to infinity $\scri$ and $\tO$, $\rO$ 
are tangent to it. The vectors $\tS$, $\qS$, $\rS$ of the algebraically 
special tetrad are drawn only in diagrams (a), (b) and (d); 
for simplicity they are omitted in the diagram (c), but see \eqref{RefSpcTimelike3}.
Different diagrams correspond to different orientations of PNDs with respect to
the infinity ${\scri}$: in the diagram (a) one PND is ingoing and one is outgoing 
(cf.\ section~\ref{ssc:timelike1}), in (b) both PNDs are outgoing 
(or ingoing, respectively, cf.\ section~\ref{ssc:timelike2}),
the diagram (c) shows the situation when one PND is tangent to ${\scri}$ (section~\ref{ssc:timelike3}),
and, finally, both PNDs are tangent in  (d) (section~\ref{ssc:timelike4}).
The relation of the reference and algebraically special tetrads in 
the generic cases (a) and (b) can be parametrized by the angle ${\THTS}$
(the angle between $\qO$ and the projection of $\kS$ to the space normal to ${\tO}$),
or by pseudospherical parameter ${\PSIS}$ (the lorenzian angle between ${\tO}$ 
and the projection of ${\kS}$ to ${\scri}$). These parameters are related by \eqref{THTPSIrel}.
In the case (a) the special tetrad can be obtained from the reference tetrad by a spatial
rotation in ${\qO\textdash\rO}$ plane by $\THTS$, cf.\ \eqref{RefSpcTimelike1}; 
in the case (b) the special tetrad is the reference tetrad boosted by rapidity ${\BTS}$ given by
${\sinh\BTS=\cot\THTS=\sinh^{\!-1\!}\PSIS}$, cf.\ \eqref{RefSpcTimelike2}.
}
\end{figure}

Now we will study the situation near a timelike conformal infinity (${\nsgn=-1}$)
when both distinct algebraic directions are not tangent to ${\scri}$, such that one of them
is outgoing and the other ingoing, ${\EPS_1\neq\EPS_{2\!s}}$.
In this case we require that the orientation ${\EPS_\refT}$ of the reference
tetrad is adjusted to ${\EPS_1}$, and that ${\tS}$ is aligned along ${\tO}$,
\begin{equation}\label{spctetrdeftimelike1}
\tO=\tS\comma \sO=\sS\comma \EPS_\refT=\EPS_1=-\EPS_{2\!s} \commae
\end{equation}
together with the adjustment condition \eqref{AdjustedReferNorm}. Again, the algebraically
special directions ${\kS}$ and ${\lS}$ are parametrized with respect to the
reference tetrad by a single parameter~${\THTS}$:
\begin{equation}\label{kltimelike1}
\begin{aligned}
\kS&={\textstyle\frac1{\sqrt2}}\,\bigl(\tO+\cos\THTS\,\qO+\sin\THTS\,\rO\bigr)\commae\\
\lS&={\textstyle\frac1{\sqrt2}}\,\bigl(\tO-\cos\THTS\,\qO-\sin\THTS\,\rO\bigr)\period
\end{aligned}
\end{equation}
The algebraically special and reference tetrads are thus related by
\begin{equation}\label{RefSpcTimelike1}
\begin{gathered}
\tS=\tO\comma
\qS=\cos\THTS\,\qO+\sin\THTS\,\rO\comma
\rS=-\sin\THTS\,\qO+\cos\THTS\,\rO\comma
\sS=\sO\commae
\end{gathered}
\end{equation}
which is a spatial rotation in ${\qO\textdash\rO}$ plane by angle ${\THTS}$,
see figure \ref{fig:ast-T}(a).

To read out the normalized projection into ${\scri}$ of the null vectors ${\kS}$, ${\lS}$
it is useful to rewrite \eqref{kltimelike1} in a different way
\begin{equation}
\begin{aligned}
\kS&={\textstyle\frac1{\sqrt2}}\cosh^{\!-1\!}\PSIS\,\bigl(\qO+\cosh\PSIS\,\tO+\sinh\PSIS\,\rO\bigr)\commae\\
\lS&={\textstyle\frac1{\sqrt2}}\cosh^{\!-1\!}\PSIS\,\bigl(-\qO+\cosh\PSIS\,\tO-\sinh\PSIS\,\rO\bigr)\commae
\end{aligned}
\end{equation}
where we have used the parameter ${\PSIS}$ instead of ${\THTS}$ related by \eqref{THTPSIrel}.
Comparing the normalized projections of ${\PND{1}=\kS}$ and ${\PND{2\!s}=\lS}$ with
\eqref{PSIPHIdef} we find that the pseudospherical parameters ${\PSI,\,\PHI,\,\EPS}$
of the algebraically special directions are ${\PSI_1=\PSIS}$, ${\PHI_1=0}$, ${\EPS_1=\EPS_\refT}$,
and ${\PSI_{2\!s}=\PSIS}$, ${\PHI_{2\!s}=\pi}$, ${\EPS_{2\!s}=-\EPS_\refT}$
respectively. The corresponding complex parameters are
\begin{equation}\label{PNDRtimelike1}
\R_1=\tanh\frac\PSIS2\comma
\R_{2\!s}=-\coth\frac\PSIS2\period
\end{equation}

Again, the transformation \eqref{RefSpcTimelike1}  can be decomposed into boost,
null rotation with ${\kG}$~fixed,
and  null rotation with ${\lG}$~fixed, given by 
${B=2\cosh\PSIS(1+\cosh\PSIS)^{-1}}$, ${L=\frac12\tanh\PSIS}$, and
${K=-\tanh(\PSIS/2)}$. Applying these transformations
to the field components we obtain
\begin{equation}\label{spcrefspinfieldtimelike1}
\fieldP{\refT}{\!2\!s} =
\frac{(2s)!}{2^s(s!)^2} \tanh^s\!\PSIS\,\fieldP{\spcT}{s}\commae
\end{equation}
and, in more detail, for gravitational and electromagnetic fields
\begin{gather}
\begin{gathered}
\WTP{\refT}{0}=\WTP{\refT}{4}=\frac32\tanh^2\!\PSIS\,\WTP{\spcT}{2}\comma
-\WTP{\refT}{1}=\WTP{\refT}{3}=\frac32\frac{\tanh\PSIS}{\cosh\PSIS}\,\WTP{\spcT}{2}\commae\\
\WTP{\refT}{2}=\Bigl(1-\frac32\tanh^2\!\PSIS\Bigr)\,\WTP{\spcT}{2}\commae
\end{gathered}\label{spcrefgrfieldtimelike1}\\
-\EMP{\refT}{0}=\EMP{\refT}{2}=\tanh\PSIS\;\EMP{\spcT}{1}\comma
\EMP{\refT}{1}=\cosh^{\!-\!1}\!\PSIS\;\EMP{\spcT}{1}\period
\label{spcrefEMfieldtimelike1}
\end{gather}

\begin{figure}[t]
\begin{center}
\includegraphics{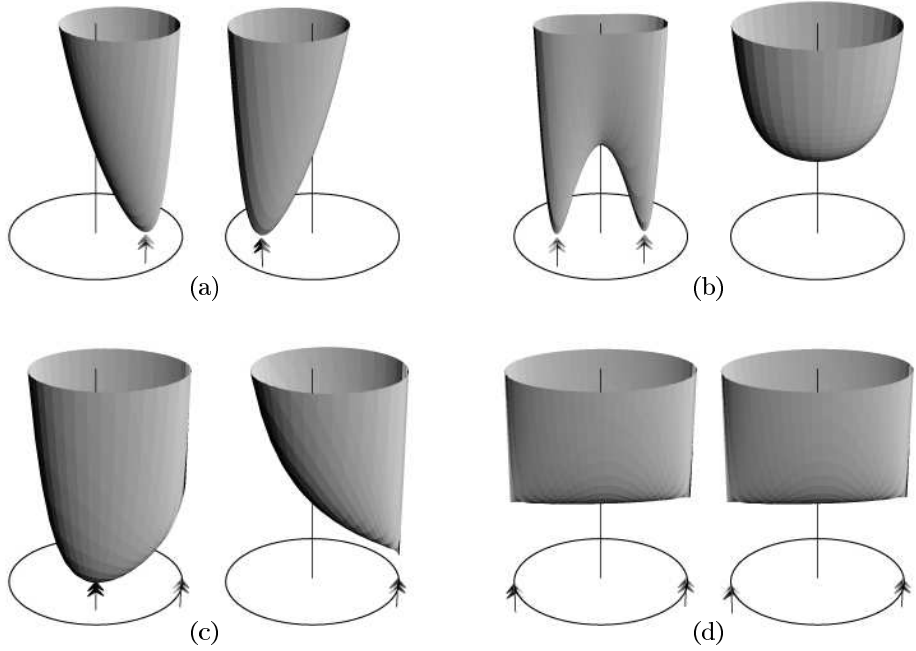}
\end{center}
\caption{\label{fig:dsr-T}%
Directional structure of gravitational radiation near a timelike infinity. 
The four diagrams, each consisting of a pair of radiation patterns, correspond to
different orientation of algebraically special 
directions with respect to the infinity ${\scri}$:
 (a) one PND  outgoing and one ingoing (cf. section~\ref{ssc:timelike1}),
 (b) both PNDs outgoing (section \ref{ssc:timelike2}; 
the case with both PNDs ingoing is analogous),
 (c) one PND tangent to ${\scri}$ and one outgoing (section \ref{ssc:timelike3}),
 (d) both PNDs tangent to ${\scri}$ (section \ref{ssc:timelike4}).
In each diagram the circles in horizontal plane represent 
a spatial projection of hemispheres of ingoing (left circle) 
and outgoing  (right circle) directions. 
The circles are parametrized by coordinates ${\RHO,\,\PHI}$ defined 
in section \ref{ssc:nulldir}, cf.\ equation~\eqref{RHOdef}.
On the vertical axis the magnitude of the radiative field component is plotted
(cf.\ \eqref{TypeDDSRgrTimelike1}, \eqref{TypeDDSRgrTimelike2}, 
\eqref{TypeDDSRgrTimelike3} and \eqref{TypeDDSRgrTimelike4}). 
The arrows  indicate mirror reflection with respect to ${\scri}$ of
the algebraically special directions (PNDs). The radiative component evaluated 
along the geodesics in these directions is (for non-tangent PNDs) asymptotically vanishing.
The radiative component diverges for unphysical geodesics tangent to $\scri$
(the border of the circles) due to fixed normalization of the null directions --- see
section~\revref{5.5}.}
\end{figure}

Substituting \eqref{PNDRtimelike1}, \vague{stared} version of
\eqref{spcrefspinfieldtimelike1}, \eqref{RPSIPHIrel}, and ${\nsgn=+1}$ into
expression \eqref{TypeDDSRspin} we obtain the asymptotic directional structure
of radiation in the form%
\begin{equation}\label{TypeDDSRspintimelike1}
\fieldP{\intT}{\!2\!s} \lteq
  \frac{\EPS^s}{\afp}\,
  \frac{(2s)!}{2^s(s!)^2}\,
  \fieldP{\spcT}{s*}\,
  \Bigl[\frac{\exp(i\PHI)}{\cosh\PSIS}\,
  \Bigl(\sinh\PSI+\EPS \EPS_\refT\sinh\PSIS\cos\PHI-i\sinh\PSIS\cosh\PSI\sin\PHI\Bigr)\Bigr]^s
  \period
\end{equation}
The direction is given by pseudospherical parameters ${\PSI}$, ${\PHI}$, ${\EPS}$,
the field is characterized by the component ${\fieldP{\spcT}{s*}}$
and by the parameter ${\PSIS}$ which fixes the orientation of the
algebraically special directions with respect to infinity~${\scri}$.
Again, only the magnitude of component $\fieldP{\intT}{\!2\!s}$ has a physical meaning,
so that
\begin{gather}
\abs{\WTP{\intT}{4}} \lteq \frac{1}{\abs{\afp}}
\,\frac32\frac{\abs{\WTP{\spcT}{2*}}}{\cosh^2\PSIS}\,
  \bigl\lvert \sinh\PSI+\EPS\EPS_\refT\sinh\PSIS\cos\PHI-i\sinh\PSIS\cosh\PSI\sin\PHI \bigr\rvert^2\commae
  \label{TypeDDSRgrTimelike1}\\
4\pi\abs{\EMS_\intT}\lteq\abs{\EMP{\intT}{2}}^2\lteq
  \frac{1}{\afp^2} \,\frac{\abs{\EMP{\spcT}{1*}}^2}{\cosh^2\PSIS}\,
  \bigl\lvert \sinh\PSI+\EPS\EPS_\refT\sinh\PSIS\cos\PHI-i\sinh\PSIS\cosh\PSI\sin\PHI \bigr\rvert^2  \period
  \label{TypeDDSRemTimelike1}
\end{gather}
This directional structure is illustrated in figure \ref{fig:dsr-T}(a).

\subsection{Timelike ${\scri}$ with non-tangent PNDs, ${\EPS_1=\EPS_{2\!s}}$}
\label{ssc:timelike2}

In the previous case we have studied the directional structure of radiation near
a timelike ${\scri}$ with two PNDs oriented in opposite directions with respect to the
infinity. Now, we will discuss the situation
when  both PNDs are outgoing, or both ingoing, ${\EPS_1=\EPS_{2\!s}}$.
The derivation of the directional structure is similar to the
previous case and we will thus sketch it only briefly.

The reference tetrad is fixed by the conditions
\begin{equation}\label{spctetrdeftimelike2}
\rO=\qS\comma \sO=\sS\comma \EPS_\refT=\EPS_1=\EPS_{2\!s} \commae
\end{equation}
together with the adjustment condition \eqref{AdjustedReferNorm}.
Therefore
\begin{equation}\label{kltimelike2}
\begin{aligned}
\kS&={\textstyle\frac1{\sqrt2}}\,\sin^{\!-\!1}\!\THTS\,\bigl(\tO+\cos\THTS\,\qO+\sin\THTS\,\rO\bigr)\\
   &={\textstyle\frac1{\sqrt2}}\sinh^{\!-1\!}\PSIS\,\bigl(\qO+\cosh\PSIS\,\tO+\sinh\PSIS\,\rO\bigr)\commae\\
\lS&={\textstyle\frac1{\sqrt2}}\,\sin^{\!-\!1}\!\THTS\,\bigl(\tO+\cos\THTS\,\qO-\sin\THTS\,\rO\bigr)\\
   &={\textstyle\frac1{\sqrt2}}\sinh^{\!-1\!}\PSIS\,\bigl(\qO+\cosh\PSIS\,\tO-\sinh\PSIS\,\rO\bigr)\commae
\end{aligned}
\end{equation}
where parameters ${\THTS}$ and ${\PSIS}$ are again related by equations \eqref{THTPSIrel}.
The algebraically special and reference tetrads are thus 
\begin{equation}\label{RefSpcTimelike2}
\begin{gathered}
\tS=\sin^{\!-\!1}\!\THTS\,\tO+\cot\THTS\,\qO\comma
\qS=\rO\comma
-\rS=\cot\THTS\,\tO+\sin^{\!-\!1}\!\THTS\,\qO\comma
\sS=\sO\commae
\end{gathered}
\end{equation}
see figure \ref{fig:ast-T}(b).
Pseudospherical parameters ${\PSI,\,\PHI,\,\EPS}$
of projections of the PNDs into ${\scri}$
are ${\PSI_1=\PSIS}$, ${\PHI_1=0}$, ${\EPS_1=\EPS_\refT}$
and ${\PSI_{2\!s}=\PSIS}$, ${\PHI_{2\!s}=\pi}$, ${\EPS_{2\!s}=\EPS_\refT}$,
respectively, i.e.,
\begin{equation}\label{PNDRtimelike2}
\R_1=\tanh\frac\PSIS2\comma
\R_{2\!s}=-\tanh\frac\PSIS2\period
\end{equation}

The transformation from the algebraically special to the reference
tetrad can be decomposed into boost (\revref{3.5}),
null rotation with ${\kG}$~fixed (\revref{3.4}),
and  null rotation with ${\lG}$~fixed (\revref{3.3}) with parameters
${B=2\tanh(\PSIS/2)}$, ${L=\frac12\coth(\PSIS/2)}$, and
${K=-\tanh(\PSIS/2)}$. For the field components we obtain
\begin{equation}\label{spcrefspinfieldtimelike2}
\fieldP{\refT}{\!2\!s} =
\frac{(2s)!}{2^s(s!)^2} \coth^s\frac\PSIS2\,\fieldP{\spcT}{s}\commae
\end{equation}
and, in more detail, for gravitational and electromagnetic fields
\begin{gather}
\WTP{\refT}{0}=\frac32\tanh^2\!\frac\PSIS2\,\WTP{\spcT}{2}\comma
\WTP{\refT}{2}=-\frac12\WTP{\spcT}{2}\comma
\WTP{\refT}{4}=\frac32\coth^2\!\frac\PSIS2\,\WTP{\spcT}{2}\comma
\WTP{\refT}{1}=\WTP{\refT}{3}=0\commae
\label{spcrefgrfieldtimelike2}\\
\EMP{\refT}{0}=-\tanh\frac\PSIS2\;\EMP{\spcT}{1}\comma
\EMP{\refT}{1}=0\comma
\EMP{\refT}{2}=\cot\frac\PSIS2\;\EMP{\spcT}{1}\period
\label{spcrefEMfieldtimelike2}
\end{gather}

Substituting into the expression \eqref{TypeDDSRspin} we finally obtain
\begin{equation}\label{TypeDDSRspintimelike2}
\begin{split}
\fieldP{\intT}{\!2\!s} &\lteq
  \frac{\EPS^s}{\afp}\,
  \frac{(2s)!}{2^s(s!)^2}\,
  \fieldP{\spcT}{s*}\\
  &\quad\times\Bigl[\frac{\exp(i\PHI)}{\sinh\PSIS}\,
  \Bigl((\cosh\PSI+\EPS\EPS_\refT\cosh\PSIS)\cos\PHI-i(\EPS\EPS_\refT+\cosh\PSIS\cosh\PSI)\sin\PHI\Bigr)\Bigr]^s
  \period
\end{split}\raisetag{39pt}
\end{equation}
A phase of this component is unphysical, its magnitude can be put into the form
\begin{equation}\label{TypeDDSRspintimelike2abs}
\abs{\fieldP{\intT}{\!2\!s}} \lteq
  \frac{1}{\abs{\afp}}\,
  \frac{(2s)!}{2^s(s!)^2}\,
  \abs{\fieldP{\spcT}{s*}}\,
  \Bigl(\sinh^{\!-\!2}\!\PSIS\,\bigl(\cosh\PSIS+\EPS\EPS_\refT\cosh\PSI\bigr)^2+\sinh^2\PSI\sin^2\PHI\Bigr)^{s/2}
  \period
\end{equation}
For gravitational and electromagnetic fields it gives
\begin{gather}
\abs{\WTP{\intT}{4}} \lteq \frac{1}{\abs{\afp}}
\,\frac32{\abs{\WTP{\spcT}{2*}}}\,
  \Bigl(\sinh^{\!-\!2}\!\PSIS\,\bigl(\cosh\PSIS+\EPS\EPS_\refT\cosh\PSI\bigr)^2+\sinh^2\PSI\sin^2\PHI\Bigr)\commae
  \label{TypeDDSRgrTimelike2}\\
4\pi\abs{\EMS_\intT}\lteq\abs{\EMP{\intT}{2}}^2\lteq
  \frac{1}{\afp^2} \,{\abs{\EMP{\spcT}{1*}}^2}\,
  \Bigl(\sinh^{\!-\!2}\!\PSIS\,\bigl(\cosh\PSIS+\EPS\EPS_\refT\cosh\PSI\bigr)^2+\sinh^2\PSI\sin^2\PHI\Bigr)  \commae
  \label{TypeDDSRemTimelike2}
\end{gather}
which is illustrated in figure \ref{fig:dsr-T}(b).

\subsection{Timelike ${\scri}$, one PND tangent to ${\scri}$}
\label{ssc:timelike3}

Until now we have concentrated on a generic orientation of algebraically special
directions with respect to the conformal infinity. In this and the next sections we
are going to study the special cases when the PNDs are \emph{tangent} to ${\scri}$.
This can occur only for timelike or null conformal infinity, the later case will be discussed in section \ref{ssc:null}.

First, we assume that only one of two distinct PNDs, say ${\PND{2\!s}}$, is
tangent to ${\scri}$. Let us note that in such a case we require normalization \eqref{klspcnorm}
only for the vector ${\kS\propto\PND{1}}$, the normalization of the other PND ${\lS}$ is fixed by the condition ${\kS\spr\lS=-1}$.
We use the PND ${\kS}$ as the vector ${\kO}$,
i.e., we define the reference tetrad by conditions
\begin{equation}\label{spctetrdeftimelike3}
\kO=\kS\comma \sO=\sS\comma \EPS_\refT=\EPS_1\commae
\end{equation}
together with the condition \eqref{AdjustedReferNorm}.
The algebraically special directions ${\kS}$ and ${\lS}$ are then given in terms of the reference tetrad as
\begin{equation}\label{kltimelike3}
\kS=\frac1{\sqrt2}\,(\tO+\qO)\comma
\lS=\sqrt2\,(\tO+\rO)\commae
\end{equation}
see figure \ref{fig:ast-T}(c).
The tetrads are related by
\begin{equation}\label{RefSpcTimelike3}
\begin{gathered}
\tS=\frac32\tO+\frac12\qO+\rO\comma
\qS=-\frac12\tO+\frac12\qO-\rO\comma
\rS=\tO+\qO+\rO\comma
\sS=\sO\period
\end{gathered}
\end{equation}
Complex parametrizations of ${\PND1}$ and ${\PND{2\!s}}$ are then
\begin{equation}\label{PNDRtimelike3}
\R_1=0\comma
\R_{2\!s}=+1\period
\end{equation}

The transformation from the algebraically special to the reference
tetrad is just the null rotation with ${\kG}$~fixed with ${L=-1}$.
Applying this transformation we obtain
\begin{equation}\label{spcrefspinfieldtimelike3}
\fieldP{\refT}{\!2\!s} =
(-1)^s \frac{(2s)!}{(s!)^2} \,\fieldP{\spcT}{s}\commae
\end{equation}
specifically for ${s=2,\,1}$,
\begin{gather}
\WTP{\refT}{0}=\WTP{\refT}{1}=0\comma
\WTP{\refT}{2}=\WTP{\spcT}{2}\comma
\WTP{\refT}{3}=-3\WTP{\spcT}{2}\comma
\WTP{\refT}{4}=6\WTP{\spcT}{2}\commae
\label{spcrefgrfieldtimelike3}\\
\EMP{\refT}{0}=0\comma
\EMP{\refT}{1}=\EMP{\spcT}{1}\comma
\EMP{\refT}{2}=-2\EMP{\spcT}{1}\period
\label{spcrefEMfieldtimelike3}
\end{gather}

Substituting into \eqref{TypeDDSRspin} we get the asymptotic directional structure
of radiation
\begin{equation}\label{TypeDDSRspintimelike3}
\fieldP{\intT}{\!2\!s} \lteq
  \frac{\EPS^s}{\afp}\,
  \frac{(2s)!}{2^s(s!)^2}\,
  \fieldP{\spcT}{s*}\,
  \bigl(\EPS\EPS_\refT+\cosh\PSI-\sinh\PSI\,\exp(i\PHI)\bigr)^s
  \commae
\end{equation}
i.e.,
\begin{gather}
\abs{\WTP{\intT}{4}} \lteq
\,3\frac{\abs{\WTP{\spcT}{2*}}}{\abs{\afp}}\,
  \bigl(\EPS\EPS_\refT+\cosh\PSI\bigr)\,\bigl( \cosh\PSI-\sinh\PSI\cos\PHI\bigr)\commae
  \label{TypeDDSRgrTimelike3}\\
4\pi\abs{\EMS_\intT}\lteq\abs{\EMP{\intT}{2}}^2\lteq
  2\frac{\abs{\EMP{\spcT}{1*}}^2}{\afp^2}\,
  \bigl(\EPS\EPS_\refT+\cosh\PSI\bigr)\,\bigl( \cosh\PSI-\sinh\PSI\cos\PHI\bigr)  \commae
  \label{TypeDDSRemTimelike3}
\end{gather}
for gravitational and electromagnetic field,
see figure \ref{fig:dsr-T}(c) and (\revref{5.37}).

\subsection{Timelike ${\scri}$, two PNDs tangent to ${\scri}$}
\label{ssc:timelike4}

Next, let both the PNDs are tangent to a timelike conformal infinity.
In such a situation there exists no natural normalization of both PNDs
analogous to the condition \eqref{klspcnorm} used above. This is related
to an ambiguity in a choice of the timelike unit vector ${\tS}$ ---
we can choose any of the (future oriented) unit vectors in the plane
${\PND{1}\textdash\PND{2\!s}}$. However, despite the fact that we cannot fix
the algebraically special tetrad uniquely, the nonvanishing component
${\fieldP{\spcT}{s}}$ is independent of this ambiguity: different choices
of the special tetrad differ only by a boost in ${\PND{1}\textdash\PND{2\!s}}$ plane,
and ${\fieldP{\spcT}{s}}$ does not change under such a boost.
In the following, we arbitrarily choose one particular algebraically
special tetrad with respect to which we define the reference tetrad.
The reference tetrad thus shares
the same ambiguity as the algebraically special tetrad.

The reference tetrad is fixed simply by conditions
\begin{equation}\label{spctetrdeftimelike4}
\tO=\tS\comma \sO=\sS\comma \EPS_\refT=\EPS_1\commae
\end{equation}
and \eqref{AdjustedReferNorm}.
PNDs ${\kS}$ and ${\lS}$ are given by
\begin{equation}\label{kltimelike4}
\kS={\textstyle\frac1{\sqrt2}}\,(\tO+\rO)\comma
\lS={\textstyle\frac1{\sqrt2}\,(\tO-\rO)}\commae
\end{equation}
so that the algebraically special and reference tetrads are related by
\begin{equation}
\tS=\tO\comma
\qS=\rO\comma
\rS=-\qO\comma
\sS=\sO\period
\end{equation}
It is just a simple spatial rotation by ${\pi/2}\,$ in ${\qO\textdash\rO}$ plane,
as illustrated in figure \ref{fig:ast-T}(d).
The complex directional parameters of ${\PND1}$ and ${\PND{2\!s}}$ are
\begin{equation}\label{PNDRtimelike4}
\R_1=+1\comma
\R_{2\!s}=-1\period
\end{equation}

The transformation  can be decomposed into boost ${B=2}$,
null rotation with ${\kG}$~fixed ${L=-1/2}$,
and  null rotation with ${\lG}$~fixed ${K=1}$.
Applying these  we obtain
\begin{equation}\label{spcrefspinfieldtimelike4}
\fieldP{\refT}{\!2\!s} =
(-1)^s\,\frac{(2s)!}{2^s(s!)^2} \,\fieldP{\spcT}{s}\commae
\end{equation}
and
\begin{gather}
\WTP{\refT}{0}=\WTP{\refT}{4}={\textstyle\frac32}\,\WTP{\spcT}{2}\comma
\WTP{\refT}{2}=-{\textstyle\frac12}\WTP{\spcT}{2}\comma
\WTP{\refT}{1}=\WTP{\refT}{3}=0\commae
\label{spcrefgrfieldtimelike4}\\
\EMP{\refT}{0}=-\EMP{\refT}{2}=\EMP{\spcT}{1}\comma
\EMP{\refT}{1}=0\period
\label{spcrefEMfieldtimelike4}
\end{gather}

Substituting into \eqref{TypeDDSRspin} we get
\begin{equation}\label{TypeDDSRspintimelike4}
\begin{gathered}
\fieldP{\intT}{\!2\!s} \lteq
  \frac{(-\EPS_\refT)^s}{\afp}\,
  \frac{(2s)!}{2^s(s!)^2}\,
  \fieldP{\spcT}{s*}\,
  \Bigl(\exp(i\PHI)\bigl(\cos\PHI-i\,\EPS\EPS_\refT \cosh\PSI\,\sin\PHI\bigr)\Bigr)^s
  \commae\\
\abs{\fieldP{\intT}{\!2\!s}} \lteq
  \frac{1}{\abs{\afp}}\,
  \frac{(2s)!}{2^s(s!)^2}\,
  \abs{\fieldP{\spcT}{s*}}\,
  \bigl(1+\sinh^2\!\PSI\,\sin^2\!\PHI\bigr)^{s/2}
  \commae
\end{gathered}
\end{equation}
which for gravitational and electromagnetic fields gives
\begin{gather}
\abs{\WTP{\intT}{4}} \lteq \frac{3}{2}\frac{1}{\abs{\afp}}\,
  \abs{\WTP{\spcT}{2*}}\,
  \bigl(1+\sinh^2\!\PSI\,\sin^2\!\PHI\bigr)
  \commae\label{TypeDDSRgrTimelike4}\\
4\pi\abs{\EMS_\intT}\lteq\abs{\EMP{\intT}{2}}^2\lteq
  \frac{1}{\afp^2} \,{\abs{\EMP{\spcT}{1*}}^2}\,
  \bigl(1+\sinh^2\!\PSI\,\sin^2\!\PHI\bigr)
  \commae\label{TypeDDSRemTimelike4}
\end{gather}
see figure \ref{fig:dsr-T}(d) and (\revref{5.36}).

\subsection{Null ${\scri}$}
\label{ssc:null}

\begin{figure}[b]
\begin{center}
\includegraphics{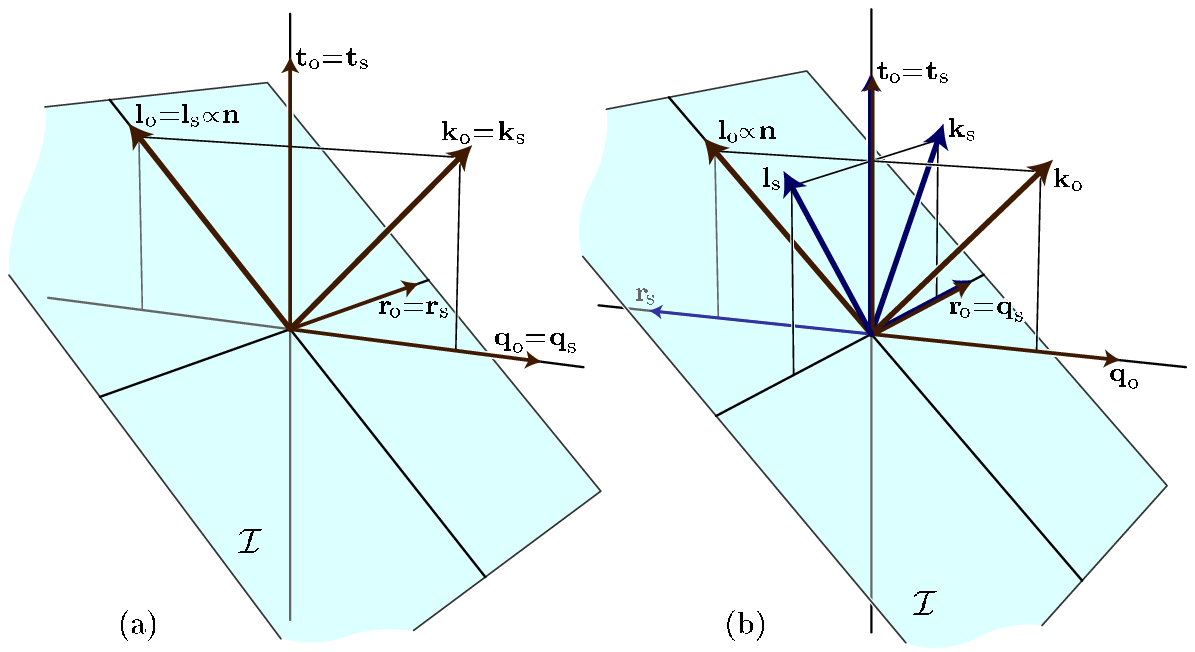}
\end{center}
\caption{\label{fig:ast-N}%
Algebraically special and reference tetrads at a null infinity. 
Vectors ${\kO,\,\lO}$ (${\kS,\,\lS}$) of the null tetrad, and
$\tO$, $\qO$, $\rO$ ($\tS$, $\qS$, $\rS$) of 
the orthonormal reference (algebraically special, respectively)
tetrad are shown; the direction ${\sO=\sS}$ 
tangent to $\scri$ and orthogonal to PNDs is not plotted.
The vectors $\kS$, $\lS$ are aligned with algebraically special directions.
The diagram (a) depicts the situation with one PND tangent to ${\scri}$.
In this case the algebraically special tetrad can be used as the reference tetrad.
In the diagram (b) neither of both distinct PNDs is tangent to ${\scri}$.
The algebraically special tetrad can be then obtained from the reference 
tetrad by a spatial rotation in ${\qO\textdash\rO}$ plane by ${\pi/2}$.} 
\end{figure}

Finally, we  investigate the case of conformal infinity ${\scri}$ of a null character, ${\nsgn=0}$.
It can be easily observed from \eqref{DSRspin}
(cf.\ section \revref{5.1} for more detail)
that the directional structure of radiation near the null infinity
is \emph{independent} of the direction along which
the infinity is approached. The only interesting question is whether the dominant
field component is vanishing or not --- in other words: whether the field is radiative or nonradiative.
It follows from the definition of PNDs that
the normalization factor ${\fieldP{\refT}{2\!s*}}$
in \eqref{DSRspin} is vanishing if and only if one of the PNDs is tangent to
${\scri}$. Thus, the tangency of PNDs to ${\scri}$
serves as the geometrical characterization of radiative/nonradiative fields.

Let us first assume that one of the PNDs, say ${\lS}$, is tangent to ${\scri}$.
As in the previous section we cannot fix the normalization of
algebraically special tetrad using the condition \eqref{klspcnorm};
the algebraically special tetrad cannot be selected uniquely.
Nevertheless, the field component ${\fieldP{\spcT}{s}}$
is still unique. If we choose one algebraically special
tetrad we can use it also as the reference tetrad --- it satisfies the
adjustment condition \eqref{AdjustedReferNorm}, cf.\ figure \ref{fig:ast-N}.
As mentioned above, the component
${\fieldP{\refT}{2\!s}}$ is then vanishing:%
\begin{equation}\label{TypeDDSRspinnulltang}
\fieldP{\intT}{\!2\!s} \lteq 0\period
\end{equation}

If both distinct PNDs are not tangent to ${\scri}$ we may normalize them by \eqref{klspcnorm}
and fix the reference tetrad by
the condition ${\tO=\tS}$, namely,
\begin{equation}\label{spctetrdefnull}
\tO=\tS\comma \sO=\sS\comma \EPS_\refT=\EPS_1\commae
\end{equation}
together with the adjustment condition \eqref{AdjustedReferNorm}, see figure \ref{fig:ast-N}.
In terms of the reference tetrad the PNDs ${\kS}$ and ${\lS}$ are given by
\begin{equation}\label{klnull}
\kS={\textstyle\frac1{\sqrt2}}\,(\tO+\rO)\comma
\lS={\textstyle\frac1{\sqrt2}}\,(\tO-\rO)\period
\end{equation}
and their complex parameters are
\begin{equation}\label{PNDRnull}
\R_1=+1\comma
\R_{2\!s}=-1\period
\end{equation}
The relation between the algebraically special and reference tetrads
is thus the same as in section \ref{ssc:timelike4}. Using \eqref{DSRspin}, ${\nsgn=0}$,
and relation \eqref{spcrefspinfieldtimelike4} we find that the radiative component has
no directional structure:
\begin{equation}\label{TypeDDSRspinnull}
\fieldP{\intT}{\!2\!s} \lteq
  \EPS_\refT^s\,
  \frac{(2s)!}{(s!)^2}\;
  \fieldP{\spcT}{s*}\;
  \frac{1}{\afp}
  \period
\end{equation}
In particular, for gravitational and electromagnetic field we obtain
\begin{gather}
\abs{\WTP{\intT}{4}} \lteq 3 \abs{\WTP{\spcT}{2*}}\,\frac{1}{\abs{\afp}}
  \commae\label{TypeDDSRgrnull}\\
4\pi\abs{\EMS_\intT}\lteq\abs{\EMP{\intT}{2}}^2\lteq
  2{\abs{\EMP{\spcT}{1*}}^2}\,\frac{1}{\afp^2}
  \period\label{TypeDDSRemnull}
\end{gather}

\section{Conclusions}
\label{sc:conclude}

We have analyzed the asymptotic directional structure of fields 
that are characterized by the existence of two distinct but equivalent 
algebraicaly special null directions. This involves a generic electromagnetic 
field (${s=1}$), the Petrov type~D gravitational fields (${s=2}$) having 
double degenerate principal null directions, and other possible fields 
of an integer spin $s$ which admit a pair of $s$-degenerate PNDs.

The structure of such fields near the conformal infinity depends on the 
specific orientation of these algebraically special directions with respect 
to $\scri$, and on the causal character of $\scri$. In case of a spacelike 
conformal infinity (${\Lambda>0}$) there is essentially only one possibility 
which is described in section~\ref{ssc:spacelike}, whereas for a timelike 
conformal infinity (${\Lambda<0}$) four different situations may occur that 
have to be discussed separately, see sections~\ref{ssc:timelike1}--~\ref{ssc:timelike4}. 
For the conformal infinity having a null character (${\Lambda=0}$) the asymptotic 
directional structure disappears: when one of the (degenerate) PNDs is tangent 
to~$\scri$ the radiative component vanishes, otherwise the radiation is present 
and it is independent of the direction along which the infinity is approached, 
cf. section~\ref{ssc:null}.

In all such cases we have introduced the privileged \vague{symmetric} reference 
tetrad which is naturally adapted to the algebraically special directions and 
to $\scri$. These are illustrated in figures~\ref{fig:ast-S}, \ref{fig:ast-T}, and
\ref{fig:ast-N}. With respect to these reference tetrads it is possible to 
characterize any null direction by standard (pseudo)spherical parameters. 
The corresponding explicit directional structure of radiation for a spacelike $\scri$ 
is presented in expression~\eqref{TypeDDSRspinspacelike}, and the four possibilities 
for a timelike $\scri$ are given by \eqref{TypeDDSRspintimelike1}, \eqref{TypeDDSRspintimelike2}, 
\eqref{TypeDDSRspintimelike3}, \eqref{TypeDDSRspintimelike4}.

These results generalize our previous study of the asymptotic directional 
structure of gravitational and electromagnetic radiation in the C-metric 
spacetimes \cite{KrtousPodolsky:2003,PodolskyOrtaggioKrtous:2003} 
to other fields which are of type D. On the other hand, 
the expressions presented here are more detailed and more explicit than 
those given in the review article \cite{KrtousPodolsky:review}. It would now be an interesting task 
to apply them on particular exact model spacetimes of type D. This may 
provide  a deeper insight into the geometric relation between the 
structure of the sources and the properties of radiation generated by them, 
as observed at spacelike or timelike conformal infinities.

\section*{Acknowledgments}
This work was supported by the grant GA\v{C}R 202/02/0735.

%\bibliographystyle{longprsty}
%\bibliography{D:/Library/TeX/bib/references}
%\end{document}

\end{document}